\documentclass[12pt,preprint]{aastex}

\shorttitle{Five Lensed Quasars from the SDSS}
\shortauthors{Inada et al.}

\begin{document}

\title{Five New High-Redshift Quasar Lenses from the Sloan Digital Sky Survey}


\author{
Naohisa Inada,\altaffilmark{1} 
Masamune Oguri,\altaffilmark{2} 
Min-Su Shin,\altaffilmark{3} 
Issha Kayo,\altaffilmark{4,5} 
Michael A. Strauss,\altaffilmark{3} 
Tomoki Morokuma,\altaffilmark{6} 
Donald P. Schneider,\altaffilmark{7}
Robert H. Becker,\altaffilmark{8,9}
Neta A. Bahcall,\altaffilmark{3} and 
Donald G. York\altaffilmark{10,11}
}

\altaffiltext{1}{Cosmic Radiation Laboratory, RIKEN, 2-1 Hirosawa, Wako,
                 Saitama 351-0198, Japan.} 
\altaffiltext{2}{Kavli Institute for Particle Astrophysics and
                 Cosmology, Stanford University, 2575 Sand Hill Road,
                 Menlo Park, CA 94025, USA.} 
\altaffiltext{3}{Princeton University Observatory, Peyton Hall,
                 Princeton, NJ 08544, USA.}  
\altaffiltext{4}{Department of Physics and Astrophysics, Nagoya
                 University, Chikusa-ku, Nagoya 464-8602, Japan.}
\altaffiltext{5}{Institute for the Physics and Mathematics of the Universe, 
                 University of Tokyo, 5-1-5 Kashiwanoha, Chiba 277-8582, Japan.}
\altaffiltext{6}{National Astronomical Observatory, 2-21-1 Osawa, Mitaka, 
                 Tokyo 181-8588, Japan.}
\altaffiltext{7}{Department of Astronomy and Astrophysics, The
                 Pennsylvania State University, 525 Davey Laboratory, 
                 University Park, PA 16802, USA.} 
\altaffiltext{8}{IGPP-LLNL, L-413, 7000 East Avenue, Livermore, CA 94550, USA.}
\altaffiltext{9}{Department of Physics, University of California at
                 Davis, 1 Shields Avenue, Davis, CA 95616, USA.}  
\altaffiltext{10}{Department of Astronomy and Astrophysics, The University 
                  of Chicago, 5640 South Ellis Avenue, Chicago, IL 60637, USA.}
\altaffiltext{11}{Enrico Fermi Institute, The University of Chicago,
                  5640 South Ellis Avenue, Chicago, IL 60637, USA.}


\begin{abstract}
We report the discovery of five gravitationally lensed quasars from the 
Sloan Digital Sky Survey (SDSS). All five systems are selected as two-image
lensed quasar candidates from a sample of high-redshift ($z>2.2$) SDSS quasars. 
We confirmed their lensing nature with additional imaging and 
spectroscopic observations. The new systems are 
SDSS~J0819+5356 (source redshift $z_s=2.237$, lens redshift $z_l=0.294$, and 
image separation $\theta=4\farcs04$), 
SDSS~J1254+2235 ($z_s=3.626$, $\theta=1\farcs56$), 
SDSS~J1258+1657 ($z_s=2.702$, $\theta=1\farcs28$), 
SDSS~J1339+1310 ($z_s=2.243$, $\theta=1\farcs69$), and 
SDSS~J1400+3134 ($z_s=3.317$, $\theta=1\farcs74$). 
We estimate the lens redshifts of the latter four systems to be 
$z_l=0.2-0.8$ from the colors and magnitudes of the lensing 
galaxies. We find that the image configurations of all systems 
are well reproduced by standard mass models. 
Although these lenses will not be included in our statistical 
sample of $z_s<2.2$ lenses, they expand the number of 
lensed quasars which can be used for high-redshift galaxy and 
quasar studies. 
\end{abstract}



\section{Introduction}

Gravitationally lensed quasars are unique astronomical and 
cosmological tools, as described in the review of \cite{kochanek06}. 
We can study the mass distributions of lensing objects from 
individual mass modeling, as well as the substructures in lensing 
objects \citep[e.g.,][]{kochanek91,mao98}. We can also investigate 
their interstellar media from dust extinctions \citep[e.g.,][]{falco99,munoz04} 
or absorption lines appearing in spectra of multiple quasar images 
\citep[e.g.,][]{curran07}. The statistics of lensed quasars 
and the measurement of time delays between lensed images are useful tools 
to constrain cosmological parameters \citep[e.g.,][]{refsdal64,turner90,fukugita90}.
In addition, lensed quasars sometimes provide opportunities to study the 
central structures of quasar host galaxies in detail 
through microlensing events \citep[e.g.,][]{richards04,poindexter08}. 

Motivated by these ideas, astronomers have searched for lensed quasars using
various methods and wavebands. Roughly 100 lensed quasars have been identified to
date \citep{kochanek06}. A number of homogeneously selected samples have been 
constructed \citep[e.g.,][]{maoz93}, allowing statistical studies to be done. 
For example, the Cosmic Lens All Sky Survey \citep[CLASS;][]{myers03,browne03} has 
created a sample of 22 lensed objects selected from $\sim$16,000 
radio sources. This sample has been used to obtain a variety of cosmological 
and astrophysical results \citep[e.g.,][]{rusin01,mitchell05,chae06}. 

The Sloan Digital Sky Survey \citep[SDSS;][]{york00} has discovered 
$\sim80,000$ spectroscopically identified quasars 
\citep{schneider07}. We are conducting a survey of lensed quasars 
selected from the large dataset of the SDSS. The survey, the SDSS Quasar 
Lens Search \citep[SQLS;][]{oguri06,oguri08a,inada08} has discovered more than 30 
lensed quasars \citep[e.g.,][and references therein]{kayo07,oguri08b}, 
making it the current largest lensed quasar survey. The SQLS also 
recovered nine previously known lensed quasars included in the SDSS footprint
\citep{walsh79,weymann80,surdej87,bade97,oscoz97,schechter98,
myers99,morgan01,magain88}. The first statistical sample of 11 SQLS lenses 
\citep{inada08} was constructed from the SDSS Data Release 3 quasar catalog 
\citep[4188 deg${}^{2}$;][]{schneider05}, and used to constrain dark energy 
\citep{oguri08a}. 

The SQLS restricts the statistical lens sample to ${z_s}<2.2$ 
because we cannot make a well-defined quasar sample for homogeneous lens 
surveys at higher redshifts. The SDSS quasars at ${z_s}>2.2$ are required 
to be point sources \citep[see][]{richards02}, and therefore they have a 
strong bias against the homogeneous lens candidate selection 
\citep{oguri06,inada08}. 
However, the SQLS candidate finding algorithm can easily be extended to locate 
higher redshift lensed quasars \citep{inada08}. Such high-redshift lensed 
quasars can be used as astronomical and cosmological tools to study (high-redshift) 
lensing galaxies \citep[e.g.,][]{kochanek00} and constrain the Hubble 
constant \citep[e.g.,][]{oguri07}. They are also useful for detailed studies of 
(lensed) high-redshift quasars. 
In this paper, we report the discoveries of five lensed quasars with high source
redshifts (${z_s}=2.237$--$3.626$). They were selected as lensed quasar candidates 
from the SDSS data, and were confirmed as lenses with the observations 
at the University of Hawaii 2.2-meter telescope (UH88), the
Astrophysical Research Consortium 3.5-meter telescope (ARC 3.5m), 
and the 3.58-meter Telescopio Nazionale Galileo (TNG 3.6m). All five 
candidates are confirmed to be double-image lensed quasars, with image 
separations of 1\farcs28--4\farcs04. 

The structure of this paper is as follows. Brief descriptions 
of the SDSS data and our lens candidate selection algorithm are presented in 
\S~\ref{sec:sdss}. We present the results of imaging and
spectroscopic observations to confirm the lensing hypotheses for 
the five objects and estimate the redshifts of the lensing galaxies
in \S~\ref{sec:observation}. We model the five lensed quasars 
in \S~\ref{sec:model} and summarize our results in \S \ref{sec:summary}. 
We use a standard cosmological model with matter density $\Omega_M=0.27$, 
cosmological constant $\Omega_\Lambda=0.73$, and Hubble constant 
$h=H_0/100{\rm km\,sec^{-1}Mpc^{-1}}=0.71$ \citep[e.g.,][]{spergel03} 
throughout this paper. 


\section{SDSS Data and Candidate Selection}\label{sec:sdss}

SDSS~J0819+5356 was selected as a lens candidate in the SDSS-I, and the other 
four lenses were selected as lens candidates in the SDSS-II Sloan Legacy Survey. 
The SDSS consists of a photometric  
\citep{gunn98} and a spectroscopic survey, and has mapped approximately
10,000 square degrees primarily in a region centered on the North Galactic 
Cap, through the SDSS-I and the subsequent SDSS-II Legacy Survey. 
The survey was conducted with a dedicated wide-field 2.5-m telescope 
\citep{gunn06} at the Apache Point Observatory in New Mexico, USA. 
The photometric survey uses five broad-band optical filters \citep[$ugriz$,][]{fukugita96}.  
The spectroscopic survey is carried out with a multi-fiber spectrograph covering 
3800{\,\AA} to 9200{\,\AA} with a resolution of $\hbox{R}\sim1800$. 
The data in each imaging observation are processed by the 
photometric pipeline \citep{lupton01}, and then the target selection pipelines 
\citep{eisenstein01,richards02,strauss02} find quasar and galaxy candidates; 
the candidates are tiled in each plate according to the algorithm of \citet{blanton03}.
The SDSS produces very 
homogeneous data with an astrometric accuracy better than about $0\farcs1$ 
rms per coordinate \citep{pier03} and photometric zeropoint accuracy better 
than about 0.02 magnitude over the entire survey area
\citep{hogg01,smith02,ivezic04,tucker06,padmanabhan08}. The SDSS is 
continuously making its data public 
\citep{stoughton02,abazajian03,abazajian04,abazajian05,
adelman06,adelman07,adelman08}. The final release (Date Release Seven) 
was made on 2008 October 31. 

The lensed quasar candidate selection algorithm of the SQLS \citep{oguri06,
inada08} is composed of two parts. One is ``morphological selection'', which 
selects candidates as extended quasars using the difference between 
the shapes of each quasar and the 
Point Spread Function (PSF) in each field. The other method is ``color selection'',
which finds quasars with objects (usually fainter) within ${\lesssim}20''$, whose colors 
are similar to the quasars. Although the selection algorithm is basically designed for
quasars with $z_s<2.2$, we can easily extend it to target higher redshift quasars 
by shifting to longer wavelength bands, as \ion{H}{1} absorption significantly
reddens colors at wavelengths shortward of the Ly$\alpha$ emission line.
\citep{schneider91,fan99,richards02}. 
For example, we can search for quasars with 
$z_s{\lesssim}3.5$ using information from the $griz$ bands rather than the $ugri$ 
bands used at lower redshifts \citep{inada08}, and $z_s{\lesssim}4.8$ using the 
$riz$ bands. We selected 
SDSS~J0819+5356, SDSS~J1258+1657, and SDSS~J1400+3134 as lens candidates by  
morphological selection with $griz$ (or $riz$ for SDSS~J1400+3134), 
and SDSS~J1254+2235 and SDSS~J1339+1310 by color selection
with $griz$. SDSS~J0819+5356 was not selected by  
color selection despite its large image separation, because of the presence 
of the bright lensing galaxy between the two stellar components. 
We note that SDSS~J0819+5356 was also selected as a possible lensed 
Ly$\alpha$ emitting galaxy by the algorithm described in \cite{shin08}
to identify strong galaxy-galaxy lenses. 

\begin{figure}
\epsscale{.3}
\plotone{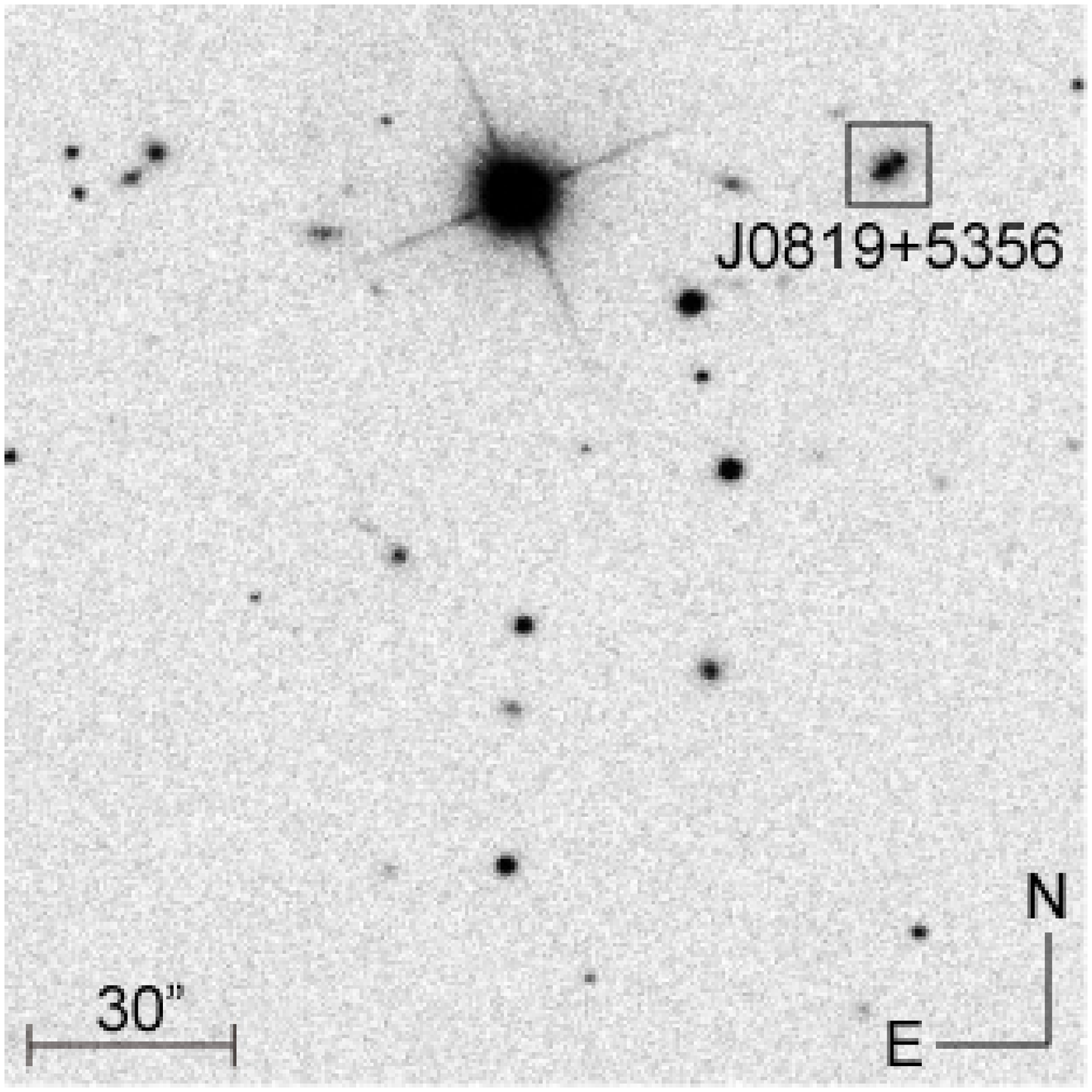}
\plotone{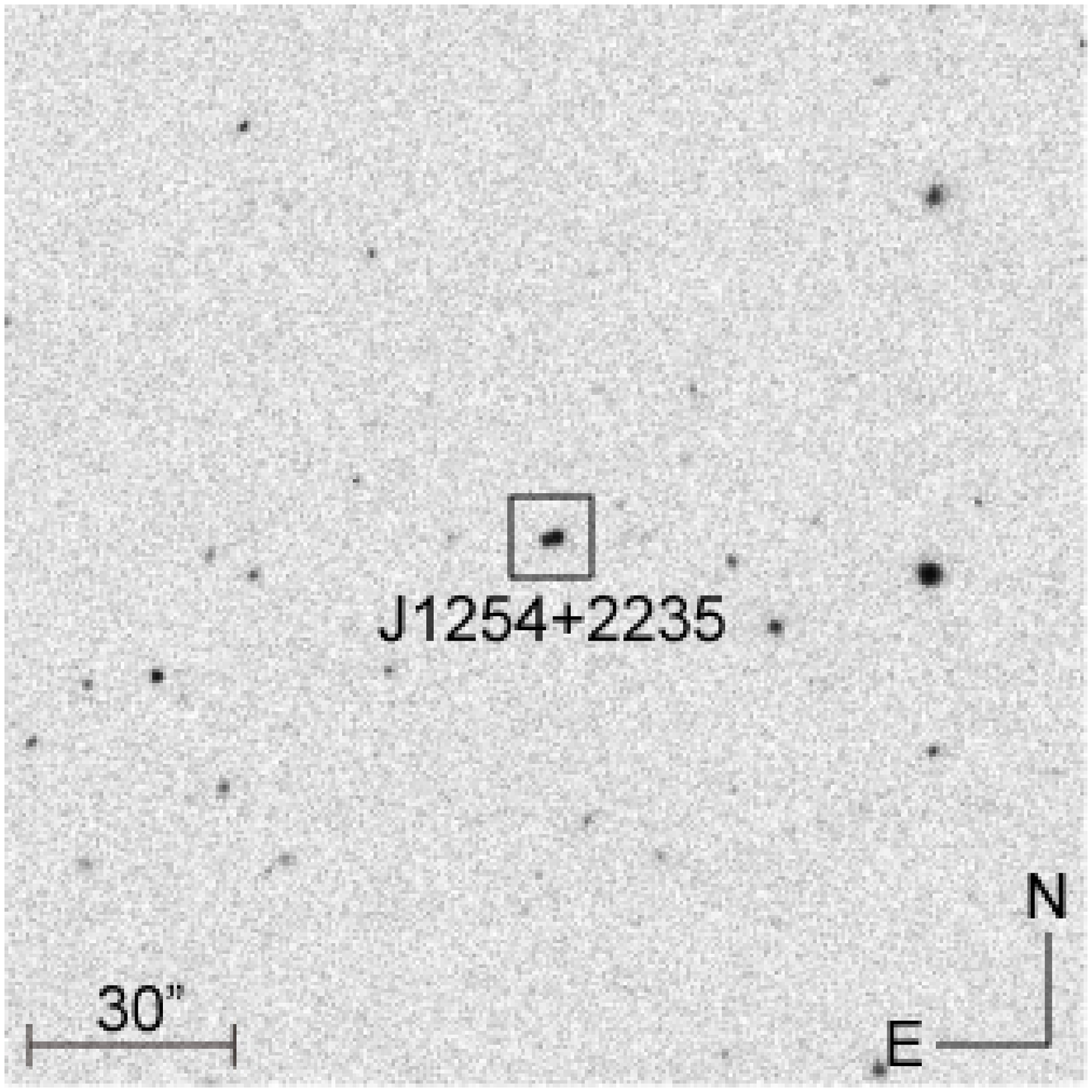}
\plotone{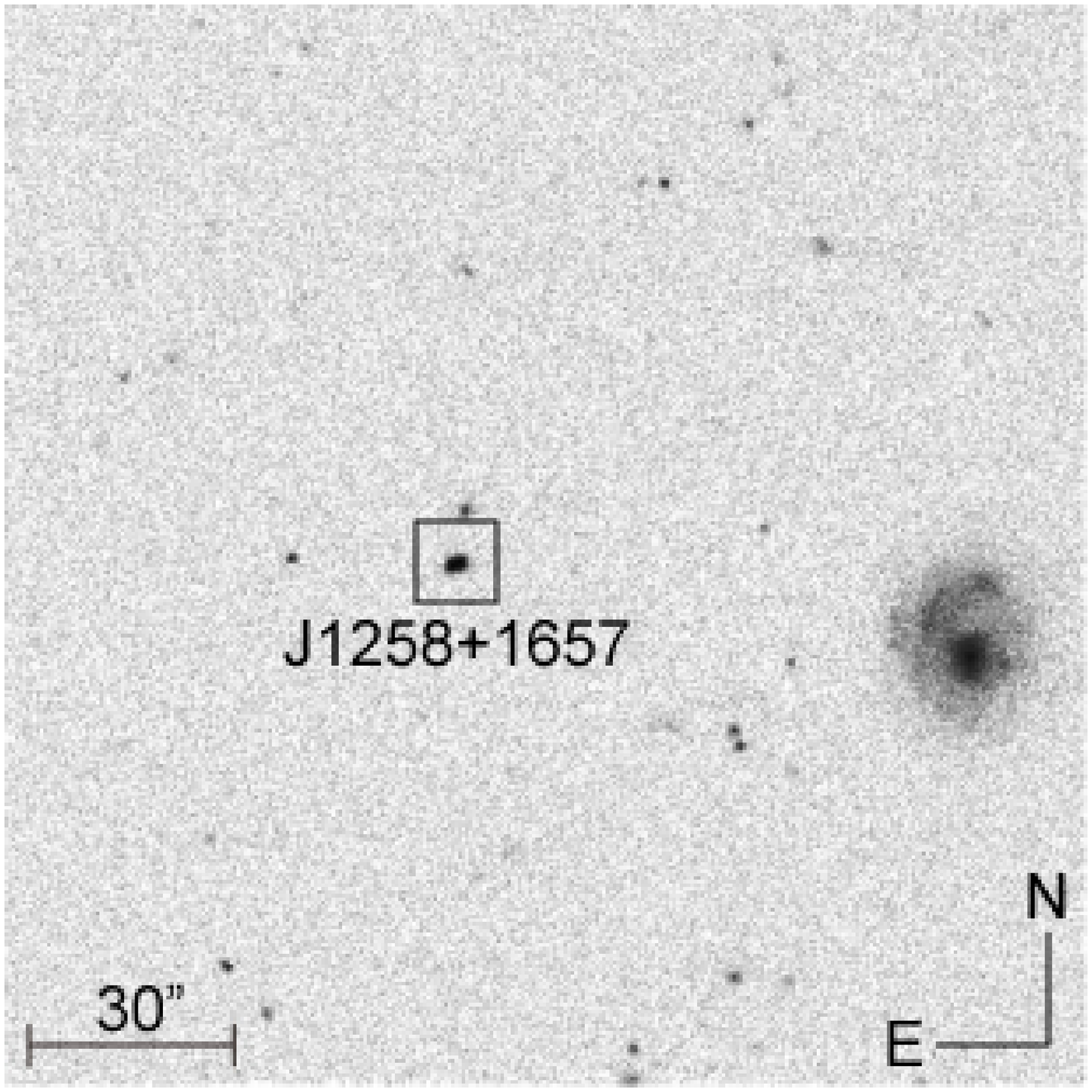}
\vspace*{0.2cm}
\plotone{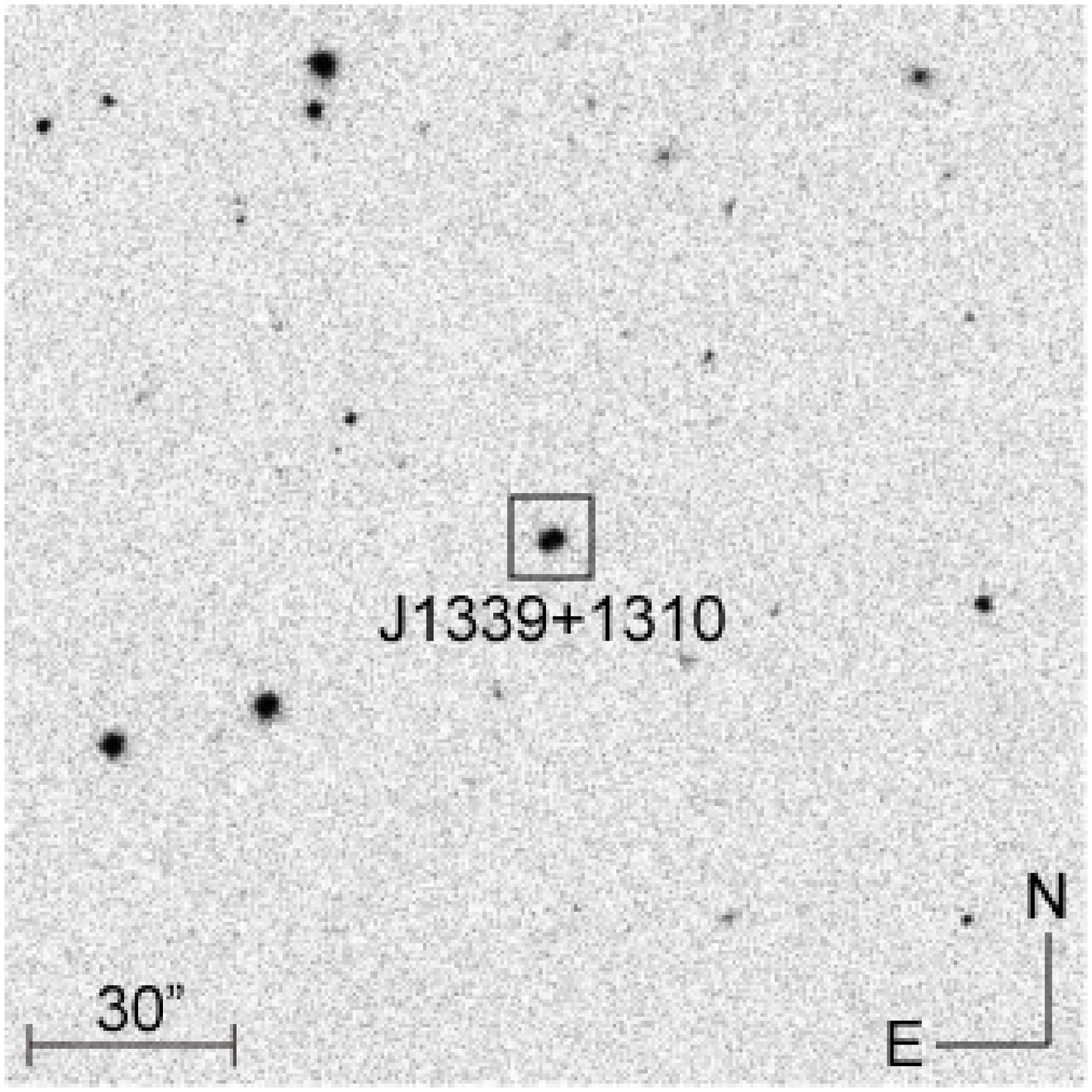}
\plotone{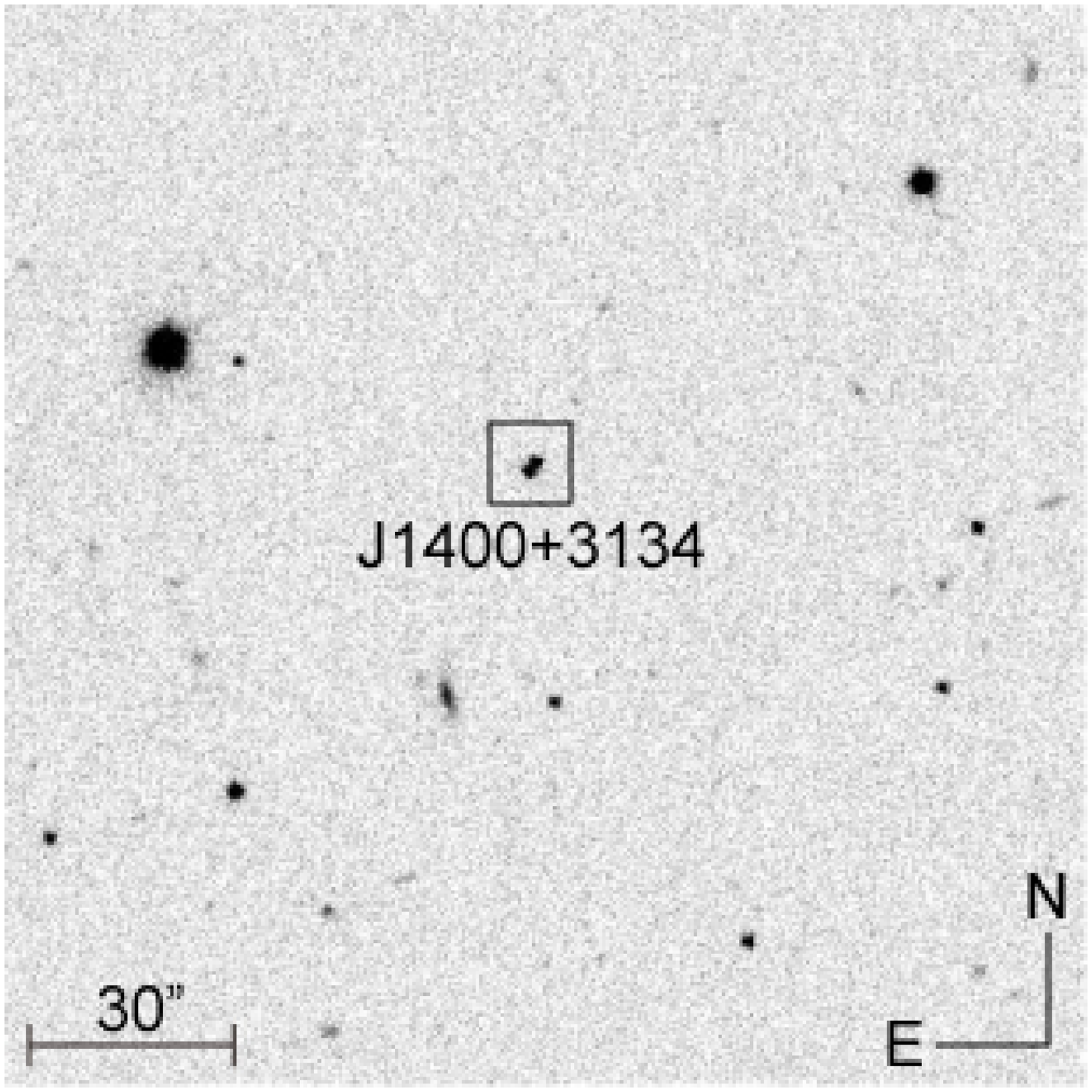}
\caption{
  Finding charts (SDSS $r$-band images) for the five lensed quasars. 
  See Table~\ref{tab:sdss} for the celestial coordinates 
  of each object. Note that SDSS~J0819+5356 is located at the edge
  of the field. The pixel scale is 0\farcs396. North is up and East is 
  to the left.
\label{fig:finding}}
\end{figure}

The SDSS $r$-band images of the fields around each lensed quasar
candidate are shown in Figure~\ref{fig:finding}. The SDSS asinh magnitudes 
\citep{lupton99} without Galactic extinction corrections and redshifts of 
the five objects are summarized in Table~\ref{tab:sdss}. The $u$-band 
asinh magnitude of SDSS J1254+2235 is not given because it is undetected in the 
$u$-band. All five candidates appear to be doubly-imaged lenses in the SDSS 
images, as we will confirm with the imaging and spectroscopic follow-up 
observations described in the next section.


\section{Observations}\label{sec:observation}

As described in \cite{inada08}, our criteria to confirm the lensing hypothesis 
for a candidate double-image lensed quasar are 1) the existence of a lensing object 
between the two stellar (quasar) components, and 2) similar spectral energy 
distributions (SEDs) for the two quasar images. 
All five candidates are marginally resolved in the SDSS imaging data and spatially 
unresolved in the SDSS spectroscopic data (the fiber diameter is 3$''$ and 
the minimum separation between each fiber on a single plate is ${\sim}55''$), and 
therefore we conducted optical/near-infrared imaging and spectroscopic follow-up 
observations to confirm their lensing natures, using the UH88, ARC 3.5m, and 
TNG 3.6m telescopes.

\subsection{Imaging Observations}

We obtained $VRI$ images for all five candidates and $B$ images for SDSS~J0819+5356
with the Tektronix 2048$\times$2048 CCD camera (Tek2k, 0\farcs22 pixel${}^{-1}$) 
at the UH88 telescope. The observations were conducted on 2007 April 11, 2007 November 
13, and 2008 March 6, with typical seeing of FWHM$\sim$0\farcs8. The exposures 
were between 300 and 480 sec depending on the magnitudes of the objects and the observing 
conditions in each night, and 800 sec for the $B$ band image of SDSS~J0819+5356.  
The instruments, observing dates, and exposure times for these  
observations are summarized in Tables~\ref{tab:followup1} and \ref{tab:followup2}. 

\begin{figure}
\epsscale{.6}
\plotone{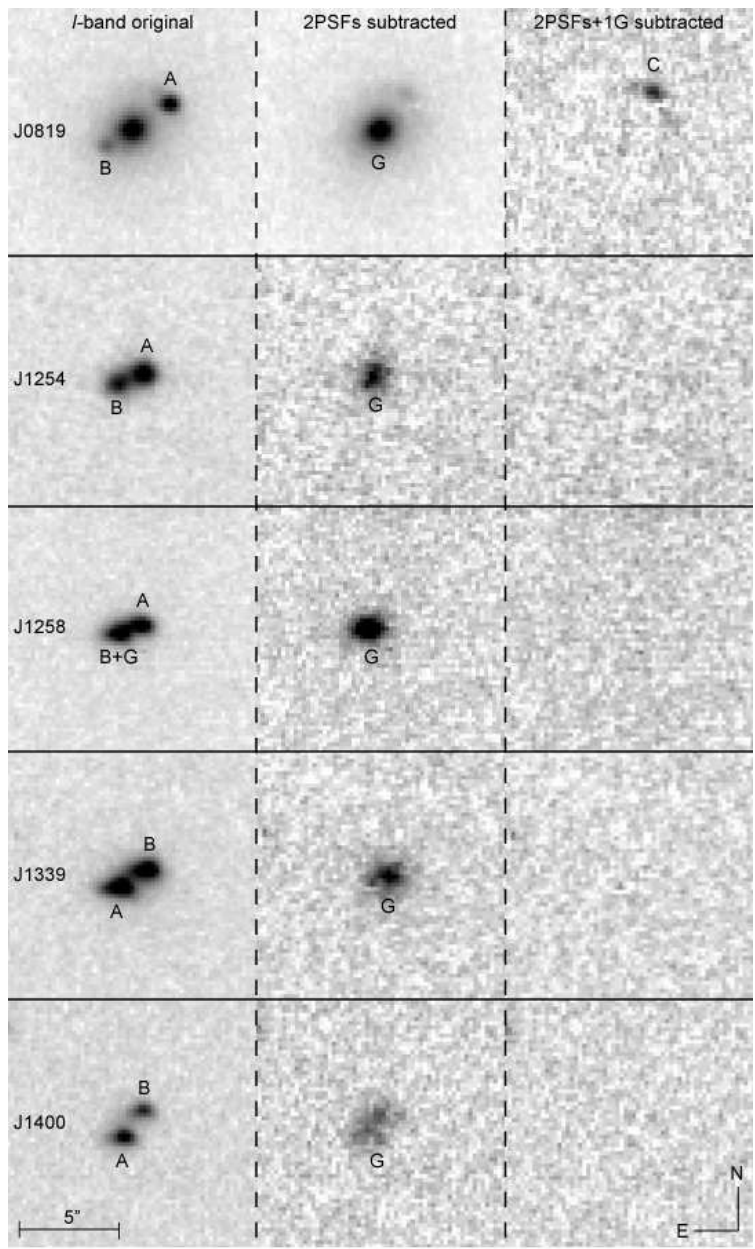}
\caption{
  The UH88 Tek2k $I$-band images of the five lensed quasars. The left 
  panels show the original image of each object. The middle panels and 
  the right panels show the residual fluxes after subtracting two PSFs,
  and two PSFs plus one galaxy component, respectively, from each original 
  image. All fits were carried out using GALFIT \citep{peng02}. 
  For SDSS~J0819+5356, the residual (component C) after subtracting 
  two PSFs plus one galaxy component may be the lensed host galaxy of 
  the source quasar. The image separations are corresponding to the 
  physical scales (at the lens redshift) of about 18 kpc, 9 kpc, 7 kpc, 
  9 kpc, and 12 kpc for SDSS~J0819+5356, SDSS~J1254+2235, SDSS~J1258+1657, 
  SDSS~J1339+1310, and SDSS~J1400+3134, respectively. 
  The image scale is 0\farcs22 ${\rm pixel^{-1}}$. 
  North is up and East is to the left. See Tables~\ref{tab:followup1} and 
  \ref{tab:followup2} for observation information. 
  \label{fig:UH-Iband}}
\end{figure}

The $I$-band images of all five candidates are shown in the left 
column of Figure~\ref{fig:UH-Iband}. In each image, we clearly detect 
two stellar components (denoted as A and B; A being the brighter component) with 
typical separations of ${\sim}1{\farcs}5$, except for SDSS~J0819+5356, which has a larger 
image separation of 4\farcs04. 
The $BVRI$ images for SDSS~J0819+5356 clearly show an 
extended object (component G) between components A and B, which we interpret as
the lensing galaxy. To see whether 
the other four candidates also have lensing galaxies between the stellar components, 
we subtracted two PSFs from the $VRI$ images of each 
candidate, using nearby stars as PSF templates. In all 12 images, the $VRI$
images of the four candidates, there is extended residual flux
between components A and B that we designate component G. 
The $I$-band PSF subtracted images are shown 
in the lower four panels of the middle column of Figure~\ref{fig:UH-Iband}. 
The morphology of the lensing galaxy of SDSS~J1400+3134 appears to be 
unusual due to its low signal-to-noise ratio. Finally 
we subtracted two PSFs plus an extended component modeled by a S\'{e}rsic profile 
using GALFIT \citep{peng02} from the $VRI$ images of the four candidates; the resulting 
images show virtually no residuals (see the lower four panels of the right column in 
Figure~\ref{fig:UH-Iband}). 
We also subtracted two PSFs, and two PSFs plus a 
galaxy component, from the $BVRI$ images of SDSS~J0819+5356. The results from the
$I$-band image are shown in the top panels of the middle and right columns of 
Figure~\ref{fig:UH-Iband}. We detect a residual flux (component C) 
around component A in the ``2PSFs+1G'' subtracted images (in all $BVRI$ 
bands). This component appears to be distorted along the critical 
curve of the system, and therefore might be the lensed host 
galaxy of the source quasar. We summarize the parameters of the 
best-fitting S\'{e}rsic profiles (in $I$-band images) of each lensing 
galaxy in Table~\ref{tab:sersic}. The very large $n$ parameter for SDSS~J0819+5356 
implies that the lensing galaxy has a steep inner profile. For 
SDSS~J1400+3134, we cannot measure the correct shape of the residual flux 
(after subtracting two PSFs) even in the $I$-band image due to its faintness.

\begin{figure}
\epsscale{.8}
\plotone{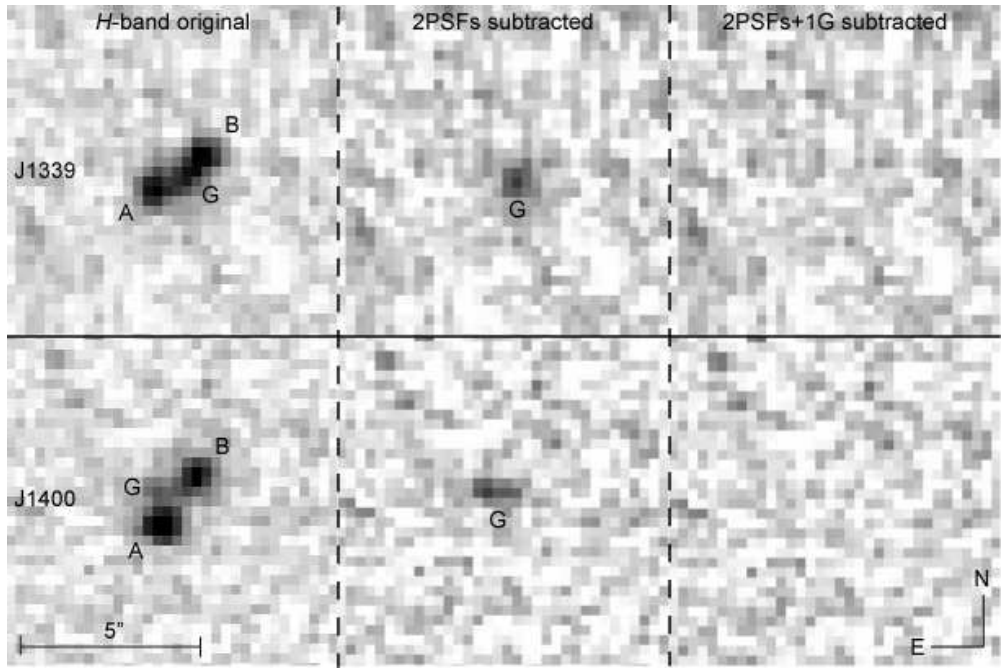}
\caption{
  The ARC 3.5m NICFPS $H$-band images of SDSS~J1339+1310 and SDSS~J1400+3134. 
  The lensing galaxies are bright in $H$-band images, and therefore we can 
  see them even in the original images. The left, middle, and right panels
  show the original images, residuals after subtracting two PSFs, and images 
  after subtracting two PSFs plus one galaxy component, respectively. The image scale 
  is 0\farcs273 ${\rm pixel^{-1}}$. North is up and East is to the left. See 
  Tables~\ref{tab:followup1} and \ref{tab:followup2} for observation information. 
  \label{fig:ARC-Hband}}
\end{figure}

Additional near-infrared ($H$-band) images for SDSS~J1339+1310 and SDSS~J1400+3134 
were taken with the Near-Infrared Camera and Fabry-Perot Spectrometer (NICFPS, 
0\farcs273 pixel${}^{-1}$) at the ARC 3.5m telescope, on 2007 March 8 and 2007 April 5. 
The exposures were 900 sec for both objects 
(see also Tables~\ref{tab:followup1} and \ref{tab:followup2}). 
The lensing galaxies are easily detected in the $H$-band images; we can see them 
even in the original images shown in the left column of Figure~\ref{fig:ARC-Hband}.
We again subtracted two PSFs and two PSFs plus a galaxy component
using GALFIT. The results 
are shown in the middle and right columns of Figure~\ref{fig:ARC-Hband}. 
The results further support the existence of the lensing objects for 
SDSS~J1339+1310 and SDSS~J1400+3134. 

The relative astrometry (from the Tek2k $I$-band images) and the absolute 
photometry \citep[Landolt-Vega system;][]{landolt92}
for the $BVRI$-band observations of the five candidates are summarized 
in Table~\ref{tab:ap}. For all candidates, the differences of the relative positions 
among each filter are less than $\sim$0\farcs05 for the stellar components and 
$\sim$0\farcs15 for the extended components. We used the standard star 
PG~0918+029 \citep{landolt92} for the optical ($BVRI$) magnitude calibration.  
We estimated the $H$ magnitudes using the 
Two Micron All Sky Survey data \citep{skrutskie06} of nearby stars. 

To summarize the imaging observations, we detect extended objects between 
the two stellar components in all five candidates, which we naturally interpret 
as the lensing galaxies. 

\subsection{Spectroscopic Observations}

To determine the SEDs of the stellar components of each candidate, we conducted 
spectroscopic observations using the Wide Field Grism Spectrograph 2 
\citep[WFGS2;][]{uehara04} at the UH88 telescope, the Dual Imaging 
Spectrograph (DIS) at the ARC 3.5m telescope, and the Device Optimized for the 
LOw RESolution (DOLORES) at the TNG 3.6m telescope. We used a 0\farcs9 long 
slit and the 300 gr/mm grism (spectral resolution of $\hbox{R}\sim600$
and spatial scale of the CCD detector of 0\farcs34 pixel${}^{-1}$) 
for WFGS2, a 1\farcs5 long slit and the B400 grism ($\hbox{R}\sim500$ and 
0\farcs40 pixel${}^{-1}$) for DIS, and, a 1\farcs0 long slit and the LR-B grism 
($\hbox{R}\sim600$ and 0\farcs252 pixel${}^{-1}$) for DOLORES. 
We aligned each slit direction to observe components A and B simultaneously. 
The exposures were typically $\sim$2000 sec for the ARC 3.5m and TNG 3.6m 
telescopes and $\sim$5000 sec for the UH88 telescope. The instruments, 
observing dates, and exposures for the spectroscopic observations are also 
summarized in Tables~\ref{tab:followup1} and \ref{tab:followup2}. 

All spectroscopic observations were conducted under good seeing conditions 
(FWHM$\lesssim$1\farcs0). The spectrum of each component 
was extracted by the standard IRAF\footnote{IRAF is distributed by 
the National Optical Astronomy Observatories, which are operated by the 
Association of Universities for Research in Astronomy, Inc., under 
cooperative agreement with the National Science Foundation.} tasks, and are shown in 
Figure~\ref{fig:spectra}. The data show that the two stellar components 
of each candidate have quite similar SEDs. In particular, the two quasar components 
of SDSS~J0819+5356 and SDSS~J1254+2235 have similar broad absorption line features. 
Therefore, together with the existence 
of the extended objects between the stellar components, we unambiguously 
conclude that SDSS~J0819+5356, SDSS~J1254+2235, SDSS~J1258+1657, SDSS~J1339+1310, 
and SDSS~J1400+3134 are all lensed quasars. 

\subsection{Lens Redshifts}

\begin{figure}
\epsscale{.32}
\plotone{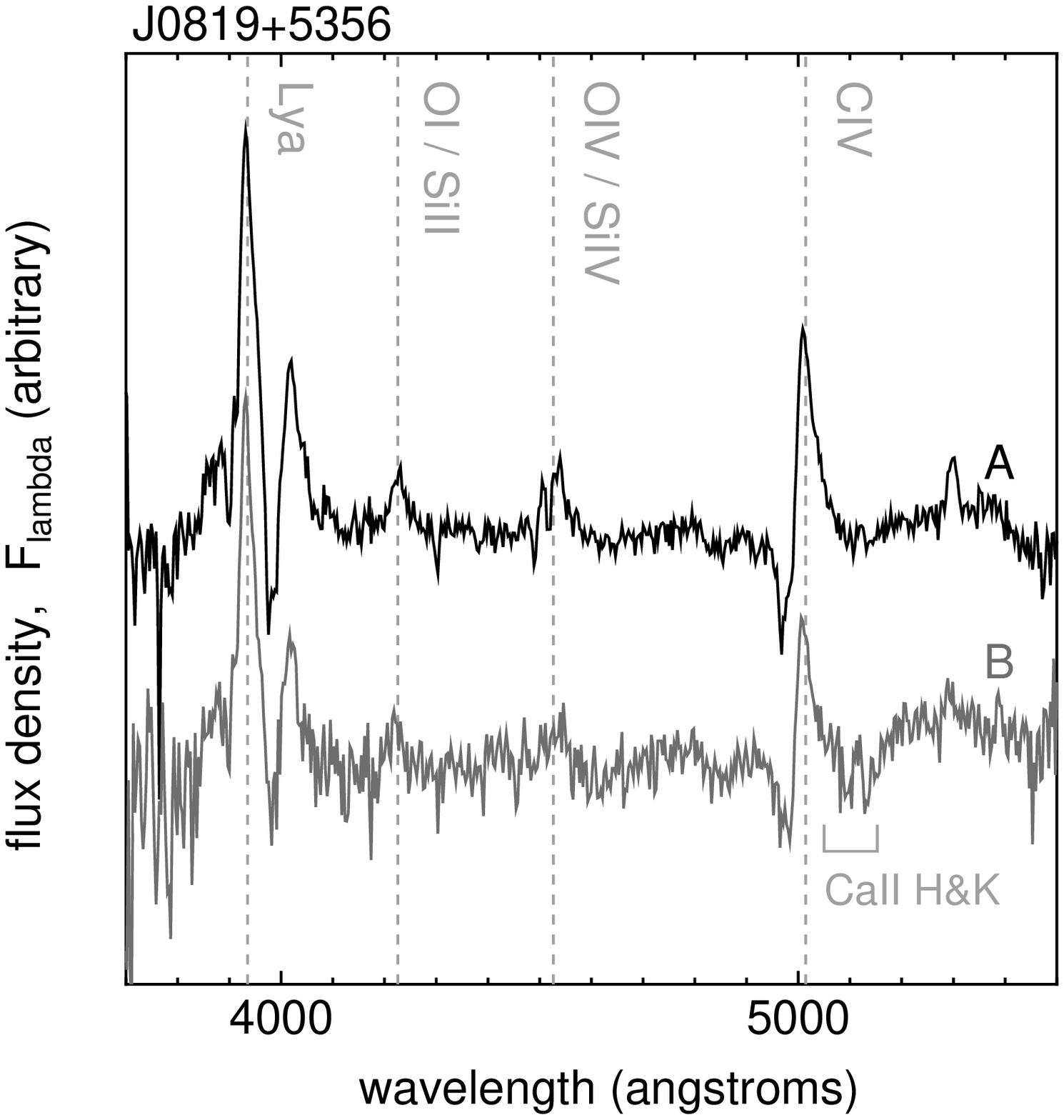}
\plotone{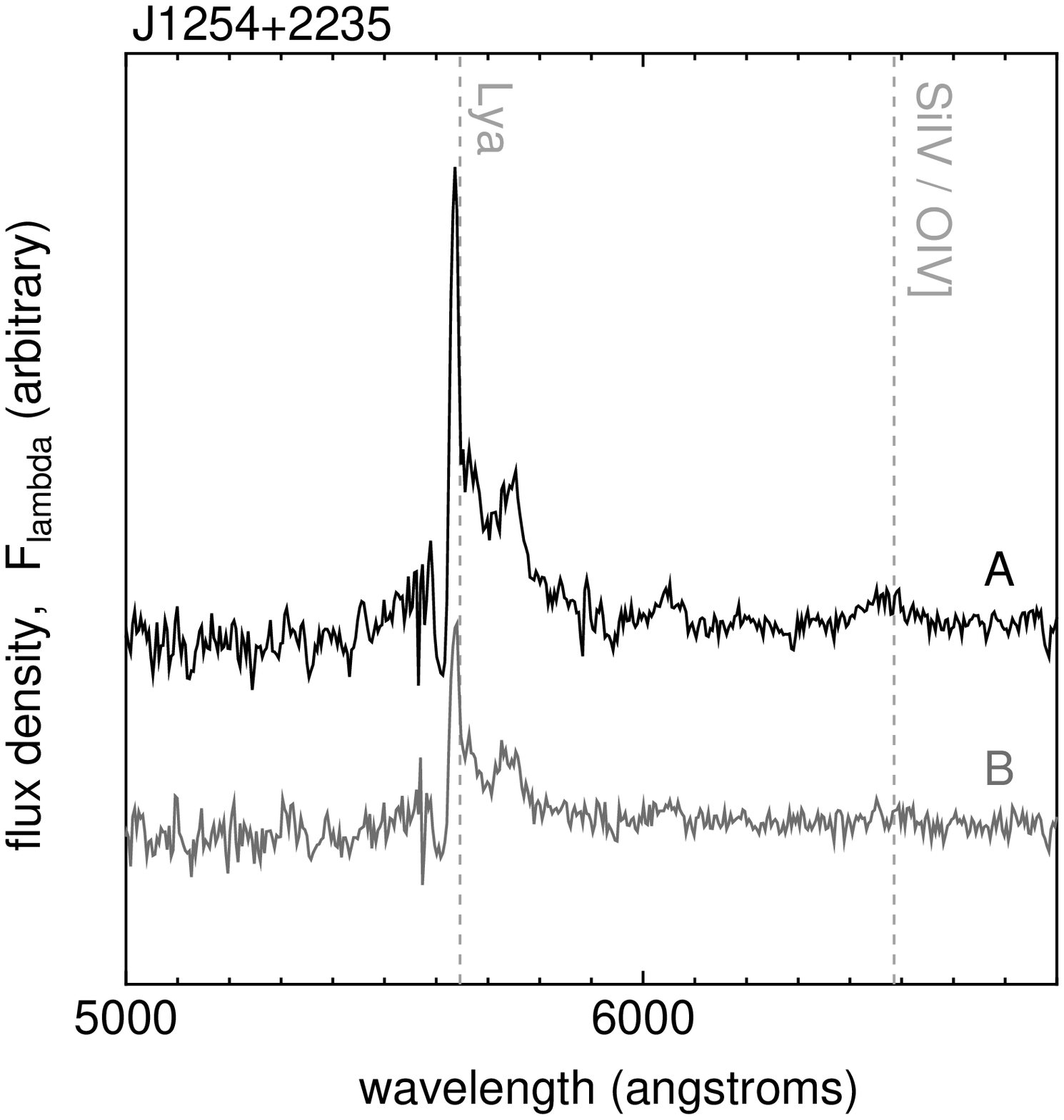}
\plotone{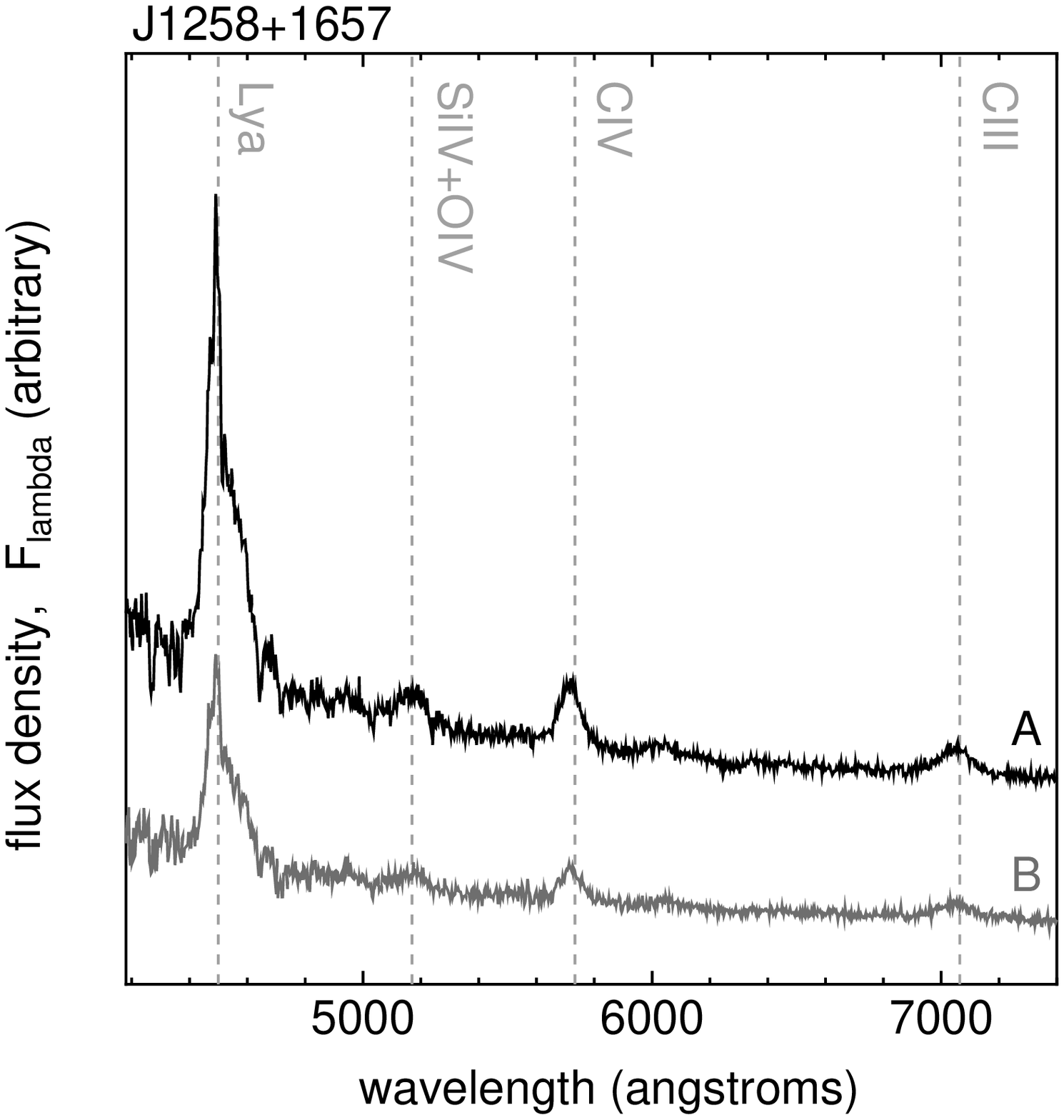}
\vspace*{0.2cm}
\plotone{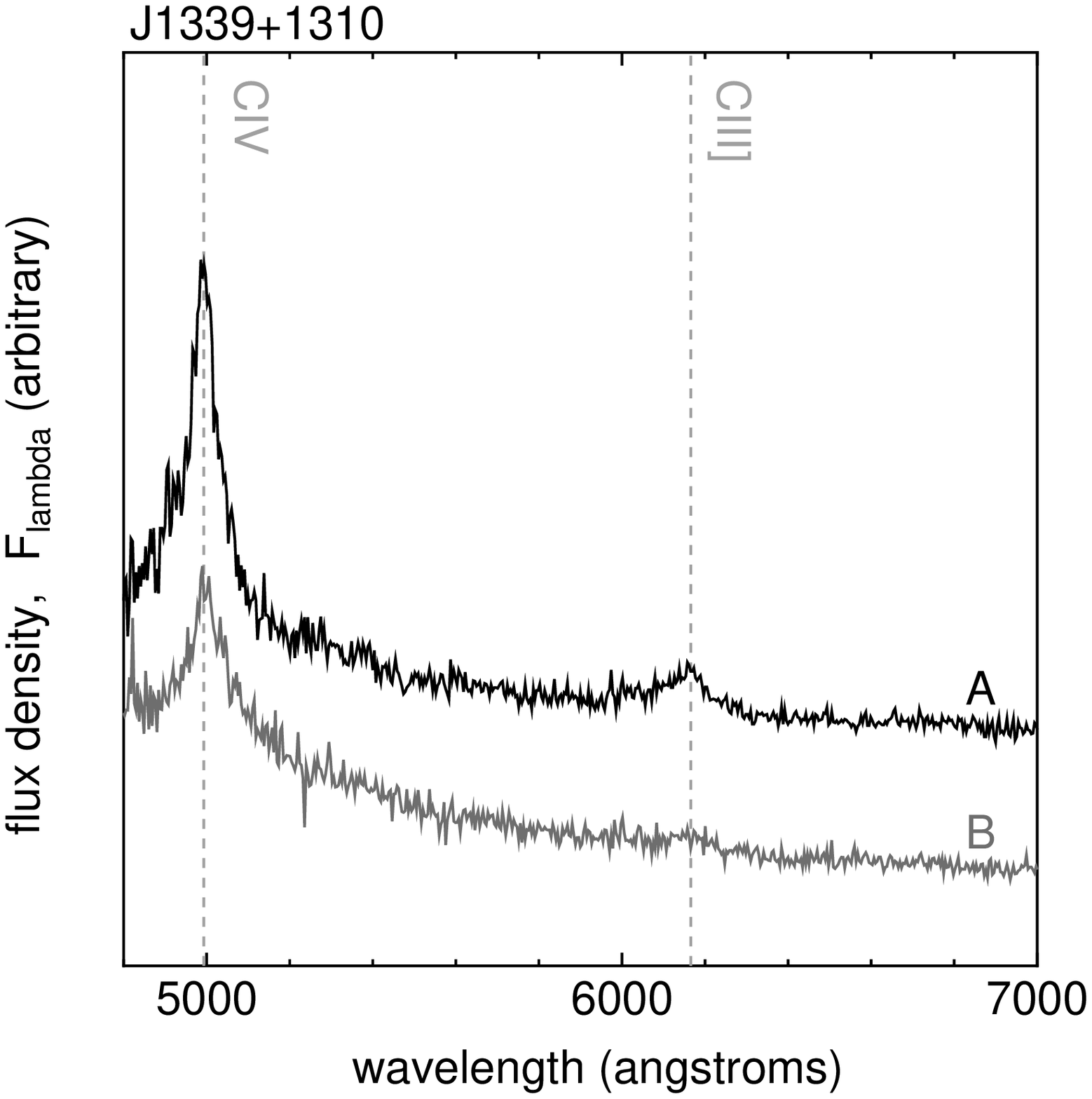}
\plotone{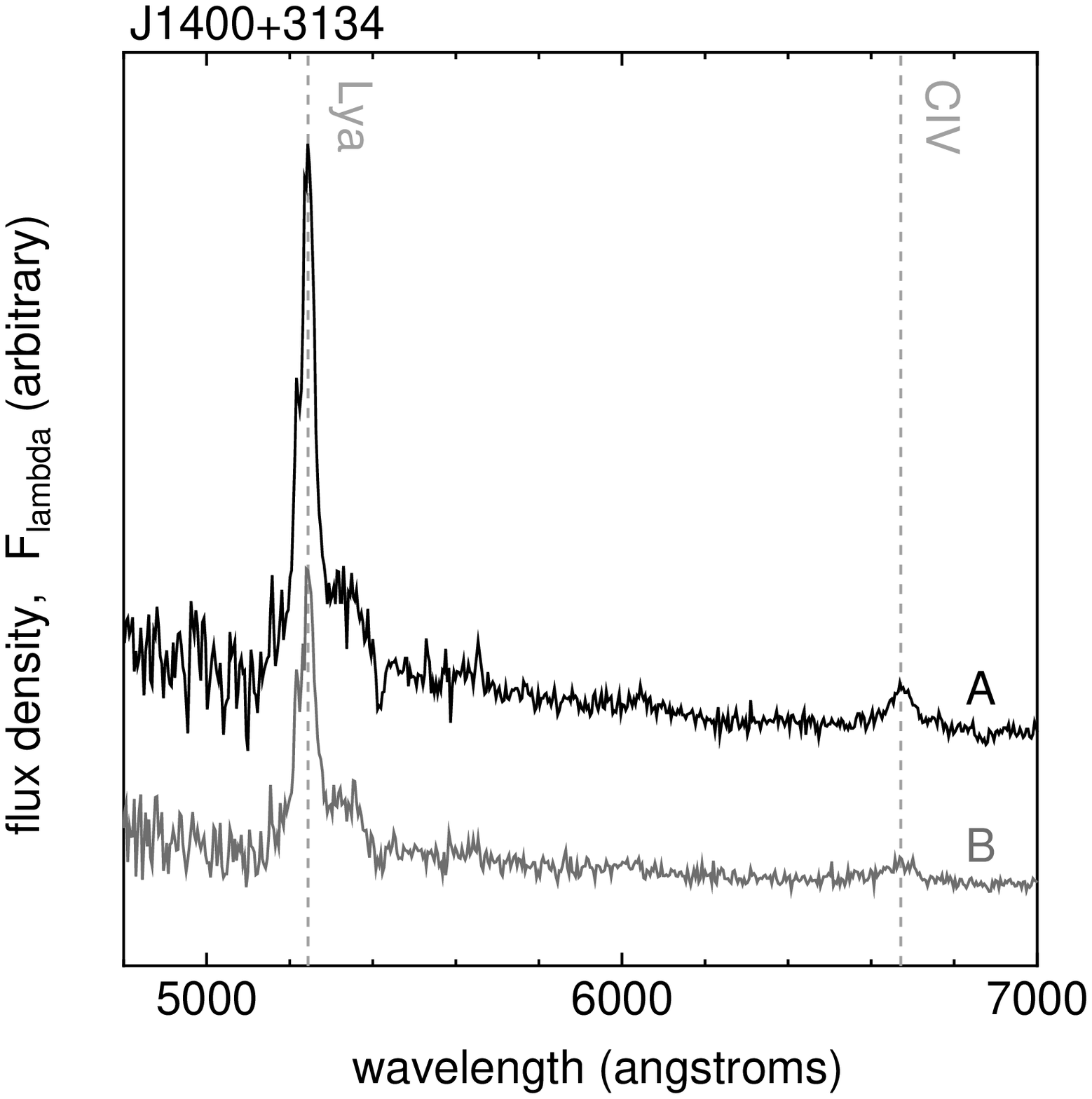}
\caption{
  Spectra of the stellar 
  components of the five lensed quasars. 
  We used the DIS at the ARC 3.5m telescope for SDSS~J0819+5356, 
  the WFGS2 at the UH88 telescope for SDSS~J1254+2235, SDSS~J1339+1310, 
  and SDSS~J1400+3134, and the DOLORES at the TNG 3.6m telescope for 
  SDSS~J1258+1657. In each panel, the spectra of brighter components 
  (component A) are shown by the black solid lines, and those of 
  fainter components (component B) are shown by the gray solid lines. 
  The vertical gray dotted lines indicate the positions of the 
  redshifted quasar emission lines. For SDSS~J0819+5356, the 
  spectra show broad absorption lines shifted shortward of 
  the \ion{N}{5} and \ion{C}{4} emission lines and the \ion{Ca}{2} 
  H\&K absorption lines from the lensing galaxy (marked by the gray
  solid symbol at $\sim$5100\AA). The two quasar components of 
  SDSS~J1254+2235 also have broad absorption line features.
  For SDSS~J1400+3134, the absorption 
  features around 5400{\AA} are real but their origin is unknown.  
  The SEDs of each pair are very similar, supporting the lensing 
  hypotheses for all objects. 
\label{fig:spectra}}
\end{figure}

\begin{figure}
\epsscale{.7}
\plotone{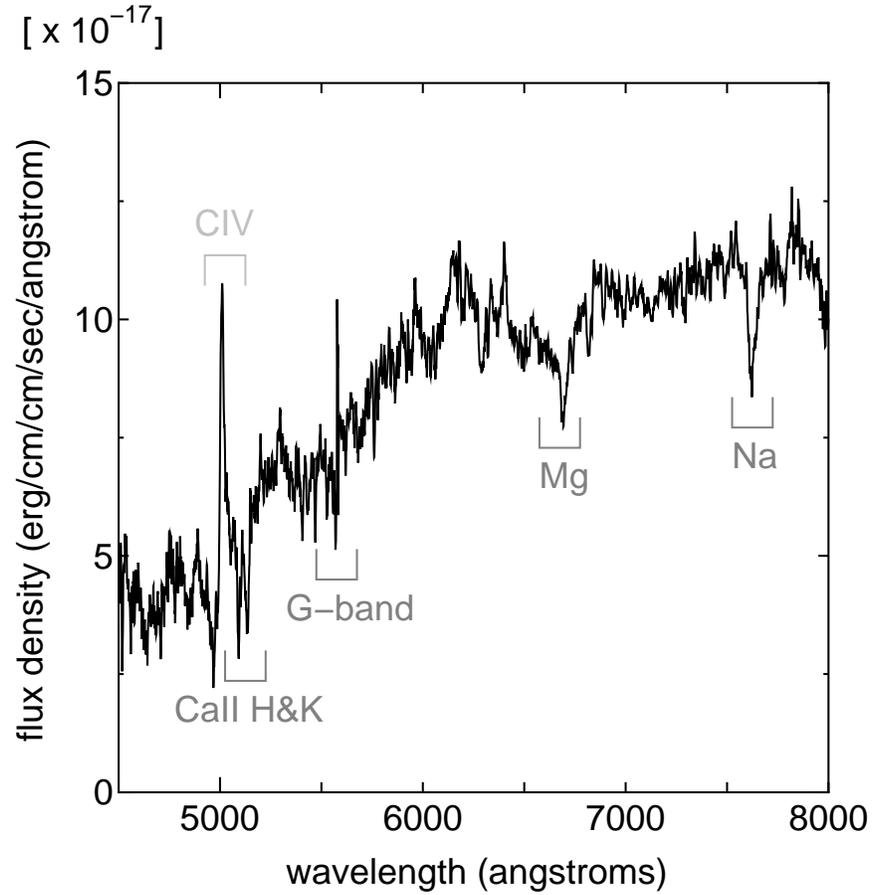}
\caption{
  The SDSS spectrum of SDSS~J0819+5356. The absorption lines from the lensing 
  galaxy at $z=0.294$ are marked by the dark gray symbols. The \ion{C}{4} emission 
  line from the source quasar at $z=2.237$ is marked by the light gray symbol.
\label{fig:j0819sdss}}
\end{figure}

Measurements of lens galaxy redshifts are important, because both 
source (quasar) and lens (galaxy) redshifts are necessary to 
convert dimensionless lensing quantities to physical units. 
Although source redshift measurements of lensed quasar systems are 
relatively easy because of prominent quasar emission lines (see Figure~\ref{fig:spectra}), 
this is not the case for direct measurements of lens galaxy redshifts 
\citep{eigenbrod06,eigenbrod07}. Indeed, we were not able to find 
any signal from the lensing galaxies in our spectra, except in SDSS~J0819+5356,  
for which we measure the redshift of the bright lensing galaxy to be
${z_l}=0.294$ from the \ion{Ca}{2} H\&K, G-band, \ion{Mg}{0}, and \ion{Na}{0} 
absorption lines\footnote{Due to these absorption lines, SDSS J0819+5356 
has two spectral classifications, ``SPEC\_QSO'' 
and ``SPEC\_GALAXY'' \citep[e.g.,][]{stoughton02}, in the SDSS data.} 
appearing in the SDSS spectrum (see Figure~\ref{fig:j0819sdss}). 
The lens redshift of SDSS~J0819+5356 is also confirmed in 
the DIS spectrum of component B, which shows the \ion{Ca}{2} H\&K lines 
at ${\sim}5300{\AA}$ (see Figure~\ref{fig:spectra}). 

For the remaining objects, we roughly estimate the redshifts of the lensing 
galaxies by comparing the observed colors with the results of \citet{fukugita95}. 
We particularly use the $R-I$ colors (Table~\ref{tab:ap}), since the lensing 
galaxies are faint in $V$-band.  
For SDSS~J1254+2235, the $R-I$ color of 0.32 indicates that the redshift of the 
lensing galaxy is not high. Therefore, combined with its S\'{e}rsic concentration 
index (Table~\ref{tab:sersic}), we estimate that the lensing galaxy is a 
late-type galaxy at ${z_l}{\sim}0.2$, rather than ${z_l}{\sim}0.5$. For 
SDSS~J1258+1657 and SDSS~J1339+1310, their S\'{e}rsic index and $R-I$ colors 
of $\sim1.0$ (Table~\ref{tab:sersic} and Table~\ref{tab:ap}) suggest that 
the lensing galaxies might be early-type galaxies at ${z_l}{\sim}0.5$. 
Assuming the early-type, we can further 
constrain the lens redshifts by comparing the observed magnitudes with the 
predicted magnitudes from the Faber-Jackson relation \citep{faber76} adopted 
by \cite{rusin03}. The predicted magnitudes ($R{\sim}20.8{\pm}0.7$ and $R{\sim}20.2{\pm}0.6$
for SDSS~J1258+1657 and SDSS~J1339+1310, respectively) assuming ${z_l}{\sim}0.4$
and using the observed image separations and Table 3 of \cite{rusin03} imply
that the lens redshifts are less than ${z_l}{\sim}0.5$ and probably ${z_l}{\sim}0.4$. 
Although the morphology of the lensing galaxy of SDSS~J1400+3134 is unknown, 
we estimate the lens redshift to be ${z_l}{\sim}0.8$ from the $R-I$ colors of 
$\sim1.5$. To summarize, we estimate the lens redshifts to be ${z_l}{\sim}0.2$, 
${z_l}{\sim}0.4$, ${z_l}{\sim}0.4$, and ${z_l}{\sim}0.8$ for SDSS~J1254+2235, 
SDSS~J1258+1657, SDSS~J1339+1310, and SDSS~J1400+3134, respectively. 

In this paper, we use these results to derive the predicted time delays of 
each system (see \S~\ref{sec:model}). These estimates should also provide a 
useful guidance for the future direct measurements of the lens redshifts. 
In particular, the direct measurement of the lens redshift of SDSS~J1339+1310 
might be easy, both because it has the relatively large image separation 
(${\theta}=1\farcs7$) and because the lensing galaxy is bright.  


\section{Mass Modeling}\label{sec:model}

We modeled the five systems using a mass model 
of a Singular Isothermal Ellipsoid (SIE). The number of the model 
parameters (8 parameters; the Einstein radius 
$R_{\rm E}$, the ellipticity $e$ and its position angle $\theta_e$, 
the position of the mass center, and the position and flux of the source 
quasar) is the same as the number of the constraints from the 
observations (8 constraints; the positions and fluxes of the two 
quasar components, and the position of the lensing galaxy), both because
all five objects are doubly-imaged lenses and because we assume the centers 
of the mass models to be the same as the galaxy light centers. 
We adopt the observables from the $I$-band images as the 
constraints, and used {\it lensmodel} \citep{keeton01} 
for modeling. As expected from the zero degree of freedom
for these models, we obtained models that fit the data perfectly 
($\chi^2{\sim}0$). The parameters of the best-fitting models and their 
1$\sigma$ (${\Delta}{\chi^2}=1$) uncertainties are summarized 
in Table~\ref{tab:model}. 

For SDSS~J0819+5356, the model ellipticity $e$ and its 
position angle $\theta_e$ (Table~\ref{tab:model}) agree with
those observed for the lensing galaxy (Table~\ref{tab:sersic}). 
Except for SDSS~J0819+5356, however, the predicted values of 
$e$ and $\theta_e$ are in poor agreement with the observed values. 
Such differences are common in lensed quasar systems when 
there are other nearby potentials producing a strong tidal shear
\citep{keeton98}. Indeed, the predicted 
$\theta_e$ of SDSS~J1258+1657 is aligned well with directions to the 
nearby two galaxies (5\farcs0 north and 7\farcs0 south from the object), 
whose photometric redshifts \citep{csabai03} of $z{\sim}0.48$ are similar 
to the estimated lens redshift. A large misalignment is also seen in 
SDSS~J1339+1310, which have some nearby galaxies with the photometric 
redshifts of $z{\sim}0.4$. We could not test the alignment for SDSS~J1400+3134, 
because we could not measure the shape of its lens galaxy in our images.

In addition to the parameters of the best-fitting models, we summarize the 
predicted time delays and total magnifications in Table~\ref{tab:model}. 
We estimated the uncertainties of the predicted time delays for the 
1$\sigma$ uncertainties of the parameters, and summarized them in Table~\ref{tab:model}. 
The measured lens redshift for SDSS~J0819+5356 and the estimated lens redshifts
for the other four lenses are used to calculate the time delays. 


\section{Summary}\label{sec:summary}

We discovered five high-redshift ($z_s>2.2$) lensed quasars, SDSS~J0819+5356, 
SDSS~J1254+2235, SDSS~J1258+1657, SDSS~J1339+1310, and SDSS~J1400+3134 from 
the SDSS. They were confirmed to be lenses by the imaging and spectroscopic 
observations at the UH88, ARC 3.5m, and TNG 3.6m telescopes. All five objects 
are two-image lensed quasars, with 
image separations of 1\farcs28--4\farcs04. The source redshifts range   
from 2.24 to 3.63. The lens redshift of SDSS~J0819+5356 is measured to be
$z_l=0.294$ from the \ion{Ca}{2} H\&K absorption lines, whereas the lens 
redshifts of the other four objects are estimated to be 0.2--0.8 from the 
colors and magnitudes of the lensing galaxies. The image configurations 
and fluxes of all the lenses are well reproduced by standard lens models. 
We find signatures of strong external shears for 
SDSS~J1258+1657 and SDSS~J1339+1310, presumably coming from nearby 
galaxies whose redshifts are estimated to be similar to that of the 
lensing galaxy. 

The statistical lensed quasar sample of the SQLS is restricted to  
$z_s<2.2$, and therefore all the lensed quasars discovered here will not 
be included in the SQLS statistical sample. The reason is that the SDSS 
quasars at ${z_s}>2.2$ are selected only from point sources and therefore 
the SDSS-selected quasars have a strong bias 
against our ``morphological selection''. Thus the five lenses will be included 
in a statistical sample when a homogeneous catalog with quasars at 
${z_s}>2.2$ is constructed. However, the five lenses will definitely 
be useful for detailed future studies, such as deep spectroscopy for 
the lensing galaxies to measure their redshifts and velocity dispersions
\footnote{Currently, measurement of velocity dispersions is probably 
possible only for the lensing galaxy of SDSS~J0819+5356.}
and for the quasar images to study the transverse 
structure in the Ly$\alpha$ forest, and high-resolution imaging to see 
the structure of the systems. In addition, monitoring observations 
to measure time delays and microlensing events will provide useful 
opportunities to study the central structures of the quasars and to constrain the 
Hubble constant. These high-redshift quasar lenses will also be important 
to extend the redshift range of the lens applications; only about 30 objects 
out of the $\sim$100 lensed quasars\footnote{CASTLES webpage 
(C.~S.~Kochanek et al., http://cfa-www.harvard.edu/castles/.)} are identified 
to be lenses at ${z_s}>2.2$.

\acknowledgments

Use of the UH 2.2-m telescope for the observations
is supported by NAOJ. Based in part on observations obtained with the Apache 
Point Observatory 3.5-meter telescope, which is owned and operated by 
the Astrophysical Research Consortium, and on observations made with the Italian 
Telescopio Nazionale Galileo (TNG) operated on the island of La Palma by the 
Fundacion Galileo Galilei of the INAF (Istituto Nazionale di Astrofisica) at 
the Spanish Observatorio del Roque de los Muchachos of the Instituto de 
Astrofisica de Canarias. N.~I. acknowledges support from the Special Postdoctoral 
Researcher Program of RIKEN and RIKEN DRI research Grant. 
This work was supported in part by Department of Energy contract
DE-AC02-76SF00515 and NSF grant AST-0707266. I.~K. acknowledges support from the 
JSPS Research Fellowships for Young Scientists and Grant-in-Aid for Scientific 
Research on Priority Areas No. 467. This work performed under the auspices of 
the U.S. Department of Energy by Lawrence Livermore National Laboratory 
under Contract DE-AC52-07NA27344.

Funding for the SDSS and SDSS-II has been provided by the Alfred P. 
Sloan Foundation, the Participating Institutions, the National Science 
Foundation, the U.S. Department of Energy, the National Aeronautics and 
Space Administration, the Japanese Monbukagakusho, the Max Planck Society, 
and the Higher Education Funding Council for England.

The SDSS is managed by the Astrophysical Research Consortium for the 
Participating Institutions. The Participating Institutions are the American 
Museum of Natural History, Astrophysical Institute Potsdam, University of Basel, 
Cambridge University, Case Western Reserve University, University of Chicago, 
Drexel University, Fermilab, the Institute for Advanced Study, the Japan 
Participation Group, Johns Hopkins University, the Joint Institute for Nuclear 
Astrophysics, the Kavli Institute for Particle Astrophysics and Cosmology, 
the Korean Scientist Group, the Chinese Academy of Sciences (LAMOST), 
Los Alamos National Laboratory, the Max-Planck-Institute for Astronomy (MPIA), 
the Max-Planck-Institute for Astrophysics (MPA), New Mexico State University, 
Ohio State University, University of Pittsburgh, University of Portsmouth, 
Princeton University, the United States Naval Observatory, and the University 
of Washington. 

This publication makes use of data products from the Two Micron All Sky Survey, 
which is a joint project of the University of Massachusetts and the Infrared 
Processing and Analysis Center/California Institute of Technology, funded by 
the National Aeronautics and Space Administration and the National Science Foundation.

{\it Facilities:} \facility{UH88 Tek2k}, \facility{UH88 WFGS2}, \facility{ARC 3.5m NICFPS}, 
\facility{ARC 3.5m DIS}, \facility{TNG 3.6m DOLORES}.



\clearpage

\clearpage

\clearpage

\clearpage

\clearpage

\clearpage

\begin{deluxetable}{ccccccccc}
\rotate
\tabletypesize{\small}
\rotate
\tablecaption{SDSS DATA OF LENSES\label{tab:sdss}}
\tablewidth{0pt}
\tablehead{ \colhead{Object} & \colhead{R.A.(J2000)} & \colhead{Decl.(J2000)} & 
\colhead{$u$} & \colhead{$g$} & \colhead{$r$} & 
\colhead{$i$} & \colhead{$z$} & 
\colhead{Redshift} 
}
\startdata
SDSS~J0819+5356 &  $124\fdg99913$  & $+53\fdg94008$  &  19.49{$\pm$}0.15 & 18.63{$\pm$}0.03 & 17.66{$\pm$}0.02 & 17.25{$\pm$}0.02 & 16.87{$\pm$}0.06 & 2.2371{$\pm$}0.0016 \\
SDSS~J1254+2235 &  $193\fdg57896$  & $+22\fdg59350$  &  \nodata\phn      & 20.30{$\pm$}0.04 & 18.95{$\pm$}0.02 & 18.68{$\pm$}0.02 & 18.32{$\pm$}0.09 & 3.6256{$\pm$}0.0019 \\
SDSS~J1258+1657 &  $194\fdg58019$  & $+16\fdg95491$  &  19.25{$\pm$}0.05 & 18.55{$\pm$}0.01 & 18.40{$\pm$}0.01 & 18.26{$\pm$}0.02 & 18.15{$\pm$}0.06 & 2.7015{$\pm$}0.0015 \\
SDSS~J1339+1310 &  $204\fdg77974$  & $+13\fdg17768$  &  18.51{$\pm$}0.03 & 18.05{$\pm$}0.01 & 18.06{$\pm$}0.01 & 17.98{$\pm$}0.02 & 17.69{$\pm$}0.05 & 2.2429{$\pm$}0.0028 \\
SDSS~J1400+3134 &  $210\fdg05322$  & $+31\fdg58170$  &  21.28{$\pm$}0.45 & 19.26{$\pm$}0.02 & 18.96{$\pm$}0.03 & 18.82{$\pm$}0.04 & 18.99{$\pm$}0.19 & 3.3166{$\pm$}0.0007 \\
\enddata
\tablecomments{Celestial coordinates (J2000), total asinh magnitudes \citep{lupton99}
without Galactic extinction correction inside aperture radii (5\farcs4, 2\farcs1, 2\farcs1, 2\farcs4, 
and 7\farcs3 for SDSS~J0819+5356, SDSS~J1254+2235, SDSS~J1258+1657, SDSS~J1339+1310, 
and SDSS~J1400+3134, respectively), and quasar emission redshifts from 
the SDSS data.} 
\end{deluxetable}

\clearpage

\begin{deluxetable}{lllll}
\tabletypesize{\scriptsize}
\rotate
\tablecaption{SUMMARY OF FOLLOWUP OBSERVATIONS 1 \label{tab:followup1}}
\tablewidth{0pt}
\tablehead{ \colhead{Object} & \colhead{Facilities for Imaging} &
\colhead{Date of Imaging} & \colhead{Facilities for Spectroscopy} & 
\colhead{Date of Spectroscopy} }
\startdata
SDSS~J0819+5356 & UH88 Tek2k ($BVRI$)                       &  2007 Nov. 13 ($B$), 2007 Apr. 11 ($VRI$) & ARC 3.5m DIS     & 2007 Oct. 20  \\
SDSS~J1254+2235 & UH88 Tek2k ($VRI$)                        &  2008 Mar. 6 ($VRI$)                      & UH88 WFGS2      & 2008 Mar. 5   \\
SDSS~J1258+1657 & UH88 Tek2k ($VRI$)                        &  2007 Apr. 11 ($VI$), 2008 Mar. 6 ($R$)   & TNG 3.6m DOLORES & 2008 Apr. 14  \\
SDSS~J1339+1310 & UH88 Tek2k ($VRI$), ARC 3.5m NICFPS ($H$)  &  2007 Apr. 11 ($VRI$), 2007 Apr. 5 ($H$)  & UH88 WFGS2      & 2007 May 13   \\
SDSS~J1400+3134 & UH88 Tek2k ($VRI$), ARC 3.5m NICFPS ($H$)  &  2007 Apr. 11 ($VRI$), 2007 Mar. 8 ($H$)  & UH88 WFGS2      & 2007 May 13   \\
\enddata
\end{deluxetable}

\clearpage

\begin{deluxetable}{ccccccc}
\tabletypesize{\small}
\rotate
\tablecaption{SUMMARY OF FOLLOWUP OBSERVATIONS 2 \label{tab:followup2}}
\tablewidth{0pt}
\tablehead{ \colhead{Object} & \colhead{Exposure ($B$)} &
\colhead{Exposure ($V$)} & \colhead{Exposure ($R$)} & 
\colhead{Exposure ($I$)} & \colhead{Exposure ($H$)} &
\colhead{Exposure (spec)} }
\startdata
SDSS~J0819+5356 & 800s        & 300s & 300s & 300s & \nodata\phn & 1800s \\
SDSS~J1254+2235 & \nodata\phn & 400s & 400s & 400s & \nodata\phn & 4800s \\
SDSS~J1258+1657 & \nodata\phn & 300s & 400s & 480s & \nodata\phn & 1200s \\
SDSS~J1339+1310 & \nodata\phn & 300s & 300s & 480s & 900s        & 3600s \\
SDSS~J1400+3134 & \nodata\phn & 300s & 300s & 400s & 900s        & 4500s \\
\enddata
\end{deluxetable}

\clearpage

\begin{deluxetable}{ccccl}
\tablewidth{0pt}
\tablecaption{PARAMETERS OF THE BEST-FITTING S\'{E}RSIC PROFILES\label{tab:sersic}}
\tablehead{\colhead{Object} & \colhead{$r_e$\tablenotemark{a}(${}''$)} &
 \colhead{$n$\tablenotemark{b}} & \colhead{$e$\tablenotemark{c}} 
 & \colhead{{$\theta_e{({}^{\circ})}$}\tablenotemark{c} }
}
\startdata
SDSS~J0819+5356  &  5.84{$\pm$}0.44  &  7.37{$\pm$}0.22  &  0.22{$\pm$}0.01  &  $-$40.33{$\pm$}0.97  \\
SDSS~J1254+2235  &  0.70{$\pm$}0.05  &  1.49{$\pm$}0.45  &  0.55{$\pm$}0.07  &  $-$18.81{$\pm$}6.55  \\
SDSS~J1258+1657  &  0.35{$\pm$}0.05  &  2.43{$\pm$}0.80  &  0.23{$\pm$}0.08  &  $-$54.42{$\pm$}17.12  \\
SDSS~J1339+1310  &  0.86{$\pm$}0.09  &  3.21{$\pm$}0.59  &  0.17{$\pm$}0.06  &  $-$18.90{$\pm$}14.41  \\
SDSS~J1400+3134  &  1.08{$\pm$}0.04  &  0.21{$\pm$}0.07  &  0.43{$\pm$}0.03  &  $-$38.57{$\pm$}3.58  \\
\enddata
\tablecomments{S\'{e}rsic parameters measured in the $I$-band images using 
GALFIT. For SDSS~J1400+3134, the shape of the lensing galaxy is not correctly
measured because of the low signal-to-noise ratio.}
\tablenotetext{a}{Effective radius of the S\'{e}rsic profile }
\tablenotetext{b}{S\'{e}rsic concentration index.}
\tablenotetext{c}{Ellipticity and its position angle. Each position angle is 
measured East of North.}
\end{deluxetable}

\clearpage

\begin{deluxetable}{crrccccc}
\tabletypesize{\footnotesize}
\tablecaption{RELATIVE ASTROMETRY AND PHOTOMETRY OF THE FIVE LENSES\label{tab:ap}}
\tablewidth{0pt}
\tablehead{\colhead{Component} & \colhead{{$\Delta$}{\rm X} (arcsec)} &
 \colhead{{$\Delta$}{\rm Y} (arcsec)} & \colhead{$B$} & \colhead{$V$} & \colhead{$R$} 
 & \colhead{$I$} & \colhead{$H$}
 }
\startdata
\multicolumn{8}{c}{SDSS~J0819+5356 ($\theta=4\farcs04$)} \vspace*{1.5mm} \\ \hline \vspace*{-2.0mm} \\ 
A &  {$\equiv$}0          & {$\equiv$}0          & 20.64{$\pm$}0.01 & 20.24{$\pm$}0.01 & 19.92{$\pm$}0.01 & 19.39{$\pm$}0.01 & \nodata\\
B &  $-$3.367{$\pm$}0.006 & $-$2.226{$\pm$}0.006 & 22.16{$\pm$}0.02 & 21.84{$\pm$}0.01 & 21.50{$\pm$}0.02 & 20.97{$\pm$}0.02 & \nodata\\
C &  $-$0.598{$\pm$}0.020 & 0.656{$\pm$}0.020    & 22.76{$\pm$}0.04 & 22.77{$\pm$}0.03 & 22.30{$\pm$}0.02 & 21.84{$\pm$}0.04 & \nodata\\
G &  $-$1.980{$\pm$}0.002 & $-$1.348{$\pm$}0.002 & 19.44{$\pm$}0.27 & 18.41{$\pm$}0.13 & 17.61{$\pm$}0.03 & 16.54{$\pm$}0.04 & \nodata\\
\cutinhead{SDSS~J1254+2235 ($\theta=1\farcs56$)}
A &  {$\equiv$}0          & {$\equiv$}0          & \nodata& 20.07{$\pm$}0.01 & 19.91{$\pm$}0.01 & 19.33{$\pm$}0.01 & \nodata\\
B &  $-$1.460{$\pm$}0.006 & $-$0.553{$\pm$}0.006 & \nodata& 20.71{$\pm$}0.04 & 20.38{$\pm$}0.01 & 19.94{$\pm$}0.04 & \nodata\\
G &  $-$0.931{$\pm$}0.036 & $-$0.259{$\pm$}0.036 & \nodata& 21.69{$\pm$}0.04 & 20.50{$\pm$}0.03 & 20.18{$\pm$}0.04 & \nodata\\
\cutinhead{SDSS~J1258+1657 ($\theta=1\farcs28$)}
A &  {$\equiv$}0          & {$\equiv$}0          & \nodata&19.34{$\pm$}0.01 & 18.99{$\pm$}0.01 & 18.58{$\pm$}0.01 & \nodata\\
B &  $-$1.183{$\pm$}0.006 & $-$0.481{$\pm$}0.006 & \nodata&19.78{$\pm$}0.03 & 19.29{$\pm$}0.04 & 19.02{$\pm$}0.04 & \nodata\\
G &  $-$0.970{$\pm$}0.035 & $-$0.274{$\pm$}0.035 & \nodata&20.67{$\pm$}0.07 & 20.38{$\pm$}0.09 & 19.23{$\pm$}0.05 & \nodata\\
\cutinhead{SDSS~J1339+1310 ($\theta=1\farcs69$)}
A &  {$\equiv$}0          & {$\equiv$}0       & \nodata&19.25{$\pm$}0.01 & 19.12{$\pm$}0.01 & 18.28{$\pm$}0.01 & 17.3{$\pm$}0.1 \\
B &  1.400{$\pm$}0.002 & 0.942{$\pm$}0.002 & \nodata&19.13{$\pm$}0.01 & 19.07{$\pm$}0.01 & 18.46{$\pm$}0.01 & 17.3{$\pm$}0.1 \\
G &  1.058{$\pm$}0.018 & 0.384{$\pm$}0.018 & \nodata&20.79{$\pm$}0.06 & 20.16{$\pm$}0.02 & 19.25{$\pm$}0.05 & 17.2{$\pm$}0.4 \\
\cutinhead{SDSS~J1400+3134 ($\theta=1\farcs74$)}
A &  {$\equiv$}0          & {$\equiv$}0       & \nodata&19.84{$\pm$}0.01 & 19.74{$\pm$}0.01 & 19.37{$\pm$}0.01 & 17.9{$\pm$}0.1 \\
B &  1.021{$\pm$}0.004 & 1.414{$\pm$}0.004 & \nodata&20.47{$\pm$}0.01 & 20.32{$\pm$}0.01 & 19.87{$\pm$}0.02 & 18.3{$\pm$}0.1  \\
G &  0.125{$\pm$}0.033 & 0.566{$\pm$}0.033 & \nodata&22.51{$\pm$}0.51 & 21.51{$\pm$}0.07 & 20.01{$\pm$}0.04 & 18.0{$\pm$}0.3 \\
\enddata
\tablecomments{Measurements are done in the Tek2k images using
 GALFIT. The positions of each component are derived in the 
 $I$-band images. The positive directions of X and Y are defined 
 by west and north, respectively.}
\end{deluxetable}

\clearpage

\begin{deluxetable}{cccccc}
\tablewidth{0pt}
\tablecaption{RESULTS OF MASS MODELING\label{tab:model}}
\tablehead{\colhead{Object} & \colhead{$R_{\rm E}$(${}''$)} &
 \colhead{$e$} &
 \colhead{$\theta_e{({}^{\circ})}$} & \colhead{$\Delta
 t$($h^{-1}$day)} & \colhead{{$\mu_{\rm tot}$} }
} 
\startdata
SDSS~J0819+5356  & 2.057{$\pm$}0.009 &  0.123{$\pm$}0.021  & $-$58.6{$\pm$}0.4   &  44.6{$\pm$}0.6   & 15.5 \\
SDSS~J1254+2235  & 0.787{$\pm$}0.008 &  0.081{$\pm$}0.044  & $-$29.2{$\pm$}12.6  &   5.2{$\pm$}0.6   &  8.5 \\
SDSS~J1258+1657  & 0.565{$\pm$}0.039 &  0.481{$\pm$}0.126  & $-$3.6{$\pm$}5.3    &  19.0{$\pm$}1.8   &  3.0 \\
SDSS~J1339+1310  & 0.853{$\pm$}0.009 &  0.287{$\pm$}0.034  & 3.0{$\pm$}2.5       &  17.7{$\pm$}1.5   &  5.6 \\
SDSS~J1400+3134  & 0.790{$\pm$}0.023 &  0.448{$\pm$}0.052  & 37.6{$\pm$}1.7      &  56.2{$\pm$}5.8   &  5.4 \\
\enddata
\tablecomments{Each position angle is measured East of North. 
In order to calculate the predicted time delays ($\Delta t$), we 
use the measured lens galaxy redshift of ${z_l}=0.294$ for SDSS~J0819+5356, 
and the estimated lens galaxy redshifts of ${z_l}=0.2$
(for SDSS~J1254+2235), ${z_l}=0.4$ (for SDSS~1258+1657 and SDSS~J1339+1310),
and ${z_l}=0.8$ (for SDSS~J1400+3134). {$\mu_{\rm tot}$} represents the 
predicted total magnification.}
\end{deluxetable}

\end{document}